# Astro2020 Science White Paper

# The Need for Laboratory Measurements and Ab Initio Studies to Aid Understanding of Exoplanetary Atmospheres

**Thematic Areas:**        <u>Planetary Systems</u>     Star and Planet Formation
     Formation and Evolution of Compact Objects         Cosmology and Fundamental Physics
     Stars and Stellar Evolution    Resolved Stellar Populations and their Environments
     Galaxy Evolution          Multi-Messenger Astronomy and Astrophysics

**Principal Author:**
Name: Jonathan Fortney
Institution: University of California, Santa Cruz
Email: jfortney@ucsc.edu
Phone:  831-502-7285

**Co-authors:** (names and institutions)
Tyler D. Robinson, Northern Arizona University
Shawn Domagal-Goldman, NASA Goddard Space Flight Center
Anthony D. Del Genio, NASA Goddard Institute for Space Studies
Iouli E. Gordon, Harvard-Smithsonian Center for Astrophysics
Ehsan Gharib-Nezhad, Arizona State University
Nikole Lewis, Cornell University
Clara Sousa-Silva, Massachusetts Institute of Technology
Vladimir Airapetian, NASA GSFC & American University
Brian Drouin, California Institute of Technology  Jet Propulsion Laboratory
Robert J. Hargreaves, Harvard-Smithsonian Center for Astrophysics
Xinchuan Huang, SETI Institute & NASA Ames Research Center
Tijs Karman, ITAMP Harvard-Smithsonian Center for Astrophysics
Ramses M. Ramirez, Tokyo Institute of Technology
Gregory B. Rieker, University of Colorado
Jonathan Tennyson, University College London
Robin Wordsworth, Harvard University
Sergei N Yurchenko, University College London
Alexandria V Johnson, Brown University
Timothy J. Lee, NASA Ames Research Center

Chuanfei Dong, Princeton University
Stephen Kane, University of California Riverside
Mercedes López-Morales, Smithsonian Astrophysical Observatory
Thomas Fauchez, NASA GSFC / USRA
Timothy Lee, NASA ARC
Mark S. Marley, NASA ARC
Keeyoon Sung, Jet Propulsion Laboratory
Nader Haghighipour, Univ. Hawaii
Tyler Robinson, Northern Arizona University
Sarah Horst, Johns Hopkins University
Peter Gao, University of California Berkeley
Der-you Kao, NASA GSFC / USRA
Courtney Dressing, University of California Berkeley
Roxana Lupu, BAER Institute / NASA AMES
Daniel Wolf Savin, Columbia University
Benjamin Fleury, Jet Propulsion Laboratory
Olivia Venot, Laboratoire Interuniversitaire des Systèmes Atmosphériques
Daniela Ascenzi, University of Trento Italy
Stefanie Milam, NASA/GSFC
Harold Linnartz, Leiden Observatory
Murthy Gudipati, Jet Propulsion Laboratory
Guillaume Gronoff, NASA LaRC/SSAI
Farid Salama, NASA ARC
Lisseth Gavilan, NASA ARC
Jordy Bouwman, Leiden Observatory
Martin Turbet, University of Geneva
Yves Benilan, Laboratoire Interuniversitaire des Systèmes Atmosphériques
Bryana Henderson, Jet Propulsion Laboratory
Natalie Batalha, University of California Santa Cruz
Rebecca Jensen-Clem, University of California Berkeley
Timothy Lyons, University of California Riverside
Richard Freedman, Seti Institute
Edward Schwieterman, University of California Riverside
Jayesh Goyal, University of Exeter
Luigi Mancini, University of Rome 2
Patrick Irwin, University of Oxford
Jean-Michel Desert, University of Amsterdam
Karan Molaverdikhani, Max Planck for Astronomy
John Gizis, University of Delaware
Jake Taylor, University of Oxford
Joshua Lothringer, University of Arizona
Raymond Pierrehumbert, University of Oxford
Robert Zellem, Jet Propulsion Laboratory
Natasha Batalha, UC Santa Cruz

Sarah Rugheimer, University of Oxford
Jacob Lustig-Yaeger, University of Washington
Renyu Hu, Jet Propulsion Laboratory
Eliza Kempton, University of Maryland
Giada Arney, NASA Goddard Space Flight Center
Mike Line, Arizona State University
Munazza Alam, Harvard-Smithsonian Center for Astrophysics
Julianne Moses, Space Science Institute
Nicolas Iro, University of Vienna
Laura Kreidberg, Harvard-Smithsonian Center for Astrophysics
Jasmina Blecic, New York University Abu Dhabi
Tom Louden, University of Warwick
Paul Mollière, Leiden Observatory
Kevin Stevenson, Space Telescope Science Institute
Mark Swain, Jet Propulsion Laboratory
Kimberly Bott, University of Washington
Nikku Madhusudhan, University of Cambridge
Joshua Krissansen-Totton, University of Washington
Drake Deming, University of Maryland
Irina Kitiashvili, NASA Ames Research Center
Evgenya Shkolnik, Arizona State University
Zafar Rustamkulov, University of California Santa Cruz
Leslie Rogers, University of Chicago
Laird Close, University of Arizona

# Introduction

We are now on a clear trajectory for improvements in exoplanet observations that will revolutionize our ability to characterize their atmospheric structure, composition, and circulation, from gas giants to rocky planets. However, exoplanet atmospheric models capable of interpreting the upcoming observations are often limited by insufficiencies in the laboratory and theoretical data that serve as critical inputs to atmospheric physical and chemical tools. Here we provide an up-to-date and condensed description of areas where laboratory and/or *ab initio* investigations could fill critical gaps in our ability to model exoplanet atmospheric opacities, clouds, and chemistry, building off a larger 2016 white paper, found at https://arxiv.org/abs/1602.06305, and endorsed by the NAS Exoplanet Science Strategy report. Now is the ideal time for progress in these areas, but this progress requires better access to, understanding of, and training in the production of spectroscopic data as well as a better insight into chemical reaction kinetics both thermal and radiation-induced at a broad range of temperatures. Given that most published efforts have emphasized relatively Earth-like conditions, we can expect significant and enlightening discoveries as emphasis moves to the exotic atmospheres of exoplanets.

## Collisional Broadening and Line Mixing Parameters

Accurate and sufficiently complete sets of spectroscopic line shape parameters are crucial to atmospheric opacity calculations to aid efforts of exoplanet atmosphere characterization and radiative transfer modeling. Among these parameters, the collisional (or pressure) broadening and line mixing, which occur due to the ability of collisions to transfer energy during absorption, result in band-wide redistributions of intensity. The collisional effects depend on absorber and broadener species absorption cross-sections and are thus molecule-pair specific, and also vary with quantum state and temperature. For many molecules, air-broadening measurements are common and are documented in the HITRAN [Gordon et al., 2017] and HITEMP [Rothman et al.,2010] databases. These databases are being extended to include additional collisional partners such as $H_2$, He, $CO_2$ and $H_2O$ [Wilzewski et al., 2016, Li et al, 2015, Delahaye et al., 2016].

Unfortunately, very little reliable information is currently available for exoplanet conditions and species. Thus, measurements and theoretical calculations on the collisional effects are needed. In addition to $N_2$ dominated atmospheres, both $CO_2$ and $H_2$ dominated atmospheres are common. In terrestrial planets, these bulk gases can determine climate states, e.g., atmospheric models of Venus, Mars and Archean Earth, require $CO_2$ collisional parameters. Due to limited availability of such data, air-broadened widths are often used with an empirical scaling factor. More exotic atmospheres are commonly hypothesized for exoplanets, only adding to the phase space for accurate modeling. The temperature regimes are also important. A majority of exoplanets found are warm sub-Neptunes (T~500-1000 K) [Batalha et al 2014]. However, the available laboratory collisional broadening data are mostly limited to terrestrial range of temperatures and theoretical data are scarce [e.g, Gamache et al, 2018]. Incorrect broadening data can lead to qualitative differences for inferences from exoplanet spectra [Gharib-Nezhad & Line, 2019]. Substantially more data is still needed to provide complete and reliable sets.

Below we give a list of absorbers (and broadeners) for which collisional parameters are required, depending on the type of planets.

- *$N_2$ and/or $O_2$ dominated:* Radiatively active species (broadeners): $H_2O$ ($N_2$, $O_2$, $H_2O$); $CO_2$ ($N_2$, $O_2$, $CO_2$); $CH_4$ ($N_2$, $O_2$, $CH_4$); $O_3$ ($N_2$, $O_2$), T/P range: 70-500 K, up to ~10 bar
- *$CO_2$ dominated:* Radiatively active species (broadeners): $H_2O$ ($CO_2$, $H_2O$); $CO_2$ ($CO_2$); $CH_4$ ($CO_2$, $CH_4$), T/P range: 70-2000 K, up to ~100 bar

- *$H_2O$ dominated:* Radiatively active species (broadeners): $H_2O$($CO_2$, $H_2O$); $CO_2$ ($CO_2$, $H_2O$); $CH_4$ ($H_2O$, $CH_4$), T/P range: 70-2000 K, up to ~100 bar
- *$H_2$ and He dominated:* Radiatively active species: $H_2O$, CO, $CO_2$, $CH_4$, $NH_3$, Na, K, Li, Rb, Cs, TiO, VO, HCN, $C_2H_2$, $H_2S$, $PH_3$; broadeners: $H_2$, He, T/P range: 70-3000 K, up to 100 bar
- *Vaporized terrestrial:* Radiatively active species: $CO_2$, $H_2O$, $SO_2$, HCl, HF, OH, CO, SiO, KOH, KCl; broadeners: $CO_2$ and $H_2O$, T/P range: 700-4000 K, up to ~100 bar

**Continuum Collision-Induced Absorption**

Collisions in dense atmospheres induce transient dipole moments in key molecules, permitting transitions that are ordinarily forbidden. This "collision-induced absorption" (CIA) generally appears as continuum-like features, underlying the monomer bands and contributes substantially to the absorption, often exceeding contributions from the monomer band. Collision-induced absorption from $H_2$, He, and H control the thermal structure and spectra of gas giants and brown dwarfs. $H_2$–$H_2$ CIA can be important for the greenhouse effect (extending the habitable zone) for super-Earths with $H_2$-enriched atmospheres [Stevenson, 1999][Pierrehumbert, 2011]. Combinations of $N_2$, $CH_4$, and $H_2$ CIA provide the primary greenhouse effect on Titan, and $CO_2$–$CO_2$ CIA is a key opacity source for Venus and other $CO_2$-rich atmospheres [Wordsworth 2010]. Our understanding of early Mars climate has evolved significantly in the past few years with new calculations and measurements of $CO_2$-$H_2$ CIA [Ramirez et al. 2014] [Wordsworth et al. 2017][Ramirez, 2017][Turbet et al. 2019]. Molecular oxygen and nitrogen CIA have very pronounced spectroscopic signatures on Earth, and could be a pressure indicator for exoplanets.

Compilations of CIA data are available from laboratory and modeling studies. Most emphasize $N_2$ and $O_2$ and temperatures relevant to Earth. Results for $H_2$ exist and have been adapted to study gas giants and brown dwarfs [Abel,2013]. Still, key gaps still exist in our knowledge, some due to a lack of opacity data or modeled opacities. A recent effort to update the HITRAN CIA database along with a wish list concerning the remaining deficiencies is presented in Ref. [Karman, 2019]. This wish list includes extending CIA data involving key bulk atmospheric constituents ($H_2$, $N_2$, $CO_2$, $O_2$) generated by collisions with relevant background gases ($H_2$, $N_2$, $CO_2$, $O_2$, $H_2O$, $CH_2$, CO, $NH_3$), and extending the temperature ranges to the broad range of conditions encountered in planetary astrophysics (i.e, 50–3,000 K).

Data from collaborations between theoretical and experimental studies would have an immediate impact on understanding of planetary habitability climate.

**Molecular Opacity Data and High Spectral Resolution**

Ground-based high-resolution near-IR spectroscopy has been increasingly used to study exoplanet atmospheres using the cross correlation method [Snellen, 2010]. However, the computation of template spectra crucially relies on the position and strength of spectral lines of the key atmospheric absorbers. For hot Jupiters these are $H_2O$, CO, $CH_4$, and $CO_2$ in the near infrared, and potentially TiO, VO and other diatomics in the optical. The cross-correlation function is very sensitive to the line positions, and therefore, utilizing inaccurate opacity data in radiative modelings strongly biases the molecular abundances [Brogi, 2019]. For spectral resolutions of 100,000 (e.g., from CRIRES), this means an accuracy >1 km s$^{-1}$ (e.g., 0.01 cm$^{-1}$ at 2.3 μm) is required.

Room temperature based laboratory measurements and databases (HITRAN, GEISA) are insufficient when applied at high temperature, where millions of additional "weak" lines become more pronounced. Radiative transfer and interpretation of spectral signatures in hydrocarbon spectra have to account for the quasi-continuum absorption [ Hargreaves et al 2015; Rey et al

2016b; Rey et al 2017; Yurchenko 2017] particularly in transparency windows difficult to quantify. Presently, CO and $H_2O$ are major species for which there are reliable line lists. However, improvements are necessary to reach the accuracy requirements above. Semi-empirical studies [Tennyson et al., 2016; Rey et al., 2016; Huang et al. 2016; Lukashevskaya et al. 2017; Fernando, 2018] are allowing substantial updates (including $H_2O$, CO, $CH_4$, $CO_2$, $N_2O$, NO, $NO_2$, $NH_3$) to be made to the HITEMP database. Experimental validation at high temperature needs to be addressed in the near future. For example, differences between line lists were highlighted in high-resolution observations of $CH_4$ at high temperature [Hargreaves et al. 2015] which will impact the ability of observers to detect this molecular via cross-correlation. Finally, improvements to the line lists of $NH_3$, $C_2H_6$, $C_2H_2$, (among many others) should be addressed for gas giants. The discovery of hot super-Earths ("vaporized terrestrial") planets is driving the requirement for line lists of so far unconsidered species [Tennyson, 2017]. Given the difficulty in obtaining high resolution spectra and opacity line lists for high temperature modeling, it is crucial to invest in the "limited" line lists [Fortenberry et al. 2014] or molecular data with "intermediate" accuracy. Reliability of the data should be accurately estimated by the theoreticians generating the data and the approximation impacts validated by modelers. Ab initio prediction have to be validated by laboratory high temperature experiments [Ghysels et al 2018; Wong et al 2019] to evaluate realistic error margins in absorption / emission. Cross-section simulations using combined theoretical and laboratory data could provide an efficient method of remotely probing the temperature of astronomical objects by comparing the relative intensity in carefully selected spectral intervals [Wong et al 2019].

**Lab Experiments on Haze Formation**

Atmospheric hazes play central roles in the dynamically, radiatively, and chemically coupled system of exoplanetary atmospheres. We need to understand plausible formation mechanisms and optical properties of haze particles to interpret observations. Understanding of the formation chemistry and thermal stability of photochemical hazes will be essential to interpreting future exoplanet spectroscopic data.

Photochemical organic aerosols are ubiquitous in the cold outer Solar System where significant $CH_4$ is present. Several photochemical-thermal equilibrium models have explored the CHON atmospheric chemistry in warm/hot exoplanets. In heavily UV irradiated atmospheres of hot Jupiters, however, ion-molecule chemistry in the ionosphere, which plays crucial roles in generation of Titan aerosols [e.g., Vuitton et al. 2019, Linden et al., 2018], is not considered yet, aside from limited studies of atomic and simple molecular ions on hot Jupiters. Laboratory investigations would provide insight on the role of coupled ion and neutral chemistry in exoplanet photochemical haze generation.

In high temperature, low pressure, heavily UV irradiated exoplanet upper atmospheres, coupled chemistry of volatile elements (CHON) and other refractory elements (S, P, alkalis, and silicate/metal vapors) could generate particulates whose chemical compositions, structures, optical properties, and thermal stabilities are poorly known. Indeed, first results towards simulating simultaneous high-temperature (~1500 K) and UV-rich (Ly-alpha) environments have shown that organic aerosols are formed under these conditions [Fleury et al 2019]. Studies under early Earth and Titan-like conditions reveal complex roles for O [Horst et al., 2018] [Gavilan et 2018] and S incorporation into haze. Other measurement needs include (1) Further kinetic studies of chemical processes in gas-to-particle conversion and heterogeneous gas-particle reaction processes, including ion-molecule and UV photochemical driven growth at temperatures covering Earth-like to hot-Jupiter-like atmospheres (~300 K to 3000 K). (2) Vapor pressures of refractory materials to estimate formation of condensate clouds. Heterogeneous condensation can occur as well, but vapor pressures of plausible gas mixtures are unknown. (3) Particle growth and loss rates, chemical and thermal stabilities under plausible reactive exoplanet environments.

**Studies of Refractory Condensate Clouds**

Refractory condensates that form in high temperature atmospheres from ~500-2000 K (e.g., magnesium silicates, corundum, iron, perovskite, at ~10 to 0.001 bar) are poorly understood for giant planets and strongly irradiated rocky planets. Understanding the process that leads to the nucleation and condensation of cloud particles depends temperature and pressure and is key to modeling the location at which clouds form. The equilibrium condensation sequence over a range of temperatures, pressures, and compositions is a point of departure for understanding more complex systems. Kinetic cloud formation models predict the formation of mixed particles made of silicates plus iron, and these pathways depend on sometimes sparse laboratory data. Lab studies of grain growth under solar nebula- like conditions exist, but there is essentially none under relevant conditions (e.g., ~1600 K and 1 bar atmosphere of $H_2$). Grains that form at lower temperatures, such as Cr, MnS, $Na_2S$, ZnS, and KCl (salts/sulfides) are also important for brown dwarfs and cooler exoplanets (Morley et al. 2012, 2013).

Issues worthy of study include the extent to which these condensates are 'pure' homogeneous vs. 'dirty' heterogeneous mixtures of multiple species, the morphology of the condensate grains, and parameters relevant to microphysical growth calculations, such as cohesion properties and growth rates. Some laboratory studies of equilibrium vapor pressures [Ferguson et al., 2004] and vapor phase nucleation of pure refractory materials [Martinez et al., 2006][Furguson, 2000] have been conducted, but again, studies conducted at relevant temperatures would most improve the current state of atmospheric modeling.

Kinetic theories, which follow grain formation around seed particles, require understanding the material properties of seed candidates, e.g., surface growth/evaporation processes. Needs include: (1) Data that allows inference of chemical pathways from the gas phase to the formation of stable condensate grains. (2) Surface reaction rates for each elementary reaction in the growth of an existing grain. (3) A complete description of nucleation processes for a broad range of conditions, including gas mixtures of varying metallicities.

**Optical Properties of Particles**

Particles (clouds, hazes, and/or aerosols) profoundly affect the radiative balance of a planet, and thus the climate and spectrum. Unfortunately, particle data are sparse and nowhere near diverse enough to account for the wide variety of atmospheric particles we anticipate on exoplanets, as past studies have focused nearly-exclusively on current Earth, cold Jupiter-like gas giants, and Titan. Future work for terrestrial planets should include sulfur-derived hazes, and for warm and hot giant planets $CH_4$ found mixed with CO, $NH_3$, $N_2$, $H_2S$, $PH_3$, yielding particulates with a wide range of compositions. Three things are needed. (1) Measurements of optical properties (scattering phase functions, single scattering albedo, etc.) of a wider diversity of particles in a greater range of atmospheric conditions (temperature, pressure, and carrier gases) from the UV to mid-IR. (2) A database in which these data are accessible by the modeling and observing communities. (3) Measurements must be made in the context of better understanding atmospheric conditions and formation pathways experienced by these species.

**Reaction Rate Constants**

Exoplanet chemical reaction rates must be known at temperatures from ~30 K to above 3000 K, and at pressures from a few microbars up to ~100 bars. Reaction rate constants for conditions not found on modern Earth are poorly constrained, and existing data typically emphasize CHON chemistry, which is limiting in the exoplanet context. There is a lack of reliable kinetic data, and in some cases thermodynamic data, for molecules with other elements, such as S, P, Si, Mg, Na, K, Ca, Al, and Ti. Thermodynamic and kinetic data for relevant heavier organic molecules and ions are also lacking. The fields of air pollution, volcanic eruption chemistry, and Titan's atmosphere have enabled study of organic haze formation reactions, but do not consider planets with different redox conditions. For early Earth, O, S, and N species into gaseous precursors and haze particles has not been well studied. For $H_2$-rich atmospheres, adding comprehensive S and P reaction mechanisms are likely the next step.

Chemistry and atmospheric dynamics are often tightly coupled. Accounting for chemical processes in General Circulation Models is critical for future characterization [Venot, 2019]. The reactions controlling transport-induced quenching of key molecular groups such as $CO$-$CH_4$-$H_2O$ and $NH_3$-$N_2$ on giant planets still needs improvement, although recent work borrowed from the combustion literature has helped considerably [Venot, 2015]. Rate-coefficient information is also needed for kinetic reactions affecting the fate of the hydrocarbon radicals on hot Jupiters. Investment in non-Earth-centric measurements and simulations of reaction rates is needed.

**UV-Driven Atmospheric Chemistry**

Atmospheres targeted for transit characterization will typically be hot (500-2500 K), with UV photochemistry influencing their disequilibrium chemistry. Photoabsorption cross-sections are usually derived from ambient or low temperature data. However, room temperature data usually underestimate exoplanet UV photoabsorption and photodissociation rates. For instance, the photoabsorption cross section of $CO_2$ was recently measured to increase by four orders of magnitude at 200 nm from 300 to 800 K [Venot, 2013][Venot, 2018]. Consequently, the lack of photoabsorption cross-sections at high temperature should be addressed more thoroughly by measuring their temperature dependence up through the VUV wavelength range (115-230 nm) for the suite of important molecules (e.g. $N_2$, $O_2$, $O_3$, $H_2O$, $CO$, $CO_2$, $CH_4$, $NH_3$, Na, K, Li, Rb, Cs, TiO, VO, HCN, $C_2H_2$, $H_2S$, $PH_3$) of planetary atmospheres.

It is also important to obtain reaction rates for important molecules by exposing them simultaneously to VUV photons (115- 230 nm) that include Ly-alpha photons at temperatures ranging from Earth-like to hot-Jupiter-like (300 - 3000 K) conditions. First studies published recently demonstrate production of CO2 and H2O from a 0.3% CO in H2 at temperatures close to 1500 K [Fleury et al., 2019]. In a coordinated effort similar to the efforts that have been made over the past decades to develop laboratory facilities that can simulate low temperatures (down to 10 K), laboratory facilities need to be developed that can handle these extremely high temperatures (up to 3000 K) that are capable of simultaneous UV-exposure and conduct quantitative spectroscopic analysis of the gas-phase composition.

**Collaboration and Sharing**

The creation, distribution and use of adequate spectroscopic data requires a well-trained and appropriately funded workforce, particularly one that includes young scientists. Stronger, better organized, long term, multilateral, community-wide collaborations and workshops should be deliberately promoted and cultivated among astronomers, modelers, experimentalists and theoreticians. It will help people understand the big picture as well as each other's specific needs, find mutual interests, provide quick feedback, etc. communication channels (such as bulletin boards, email list, data sharing approaches) should be established to keep people updated with the current status and latest progress in each field to ensure state-of-the-art methods are brought to bare on key problems.